# Nano light-source generation by electron beam irradiation of $CsPbBr_3$/$Cs_4PbBr_6$ composites

*Hikaru Saito*[*,1,2], *Kotaro Kihara*[3,4], *Midori Ikeuchi*[1], *Sotatsu Yanagimoto*[2], *Tetsuya Kubota*[2],

*Keiichirou Akiba*[2,5] *and Takumi Sannomiya*[2]

[1] Institute for Materials Chemistry and Engineering, Kyushu University, 6-1 Kasugakoen, Kasuga, Fukuoka 816-8580, Japan
[2] Department of Materials Science and Engineering, School of Materials and Chemical Technology, Institute of Science Tokyo, 4259 Nagatsuta-cho, Midori-ku, Yokohama, Kanagawa, 226-8503, Japan
[3] Department of Energy Science and Engineering, School of Engineering, Kyushu University, 6-1 Kasugakoen, Kasuga, Fukuoka 816-8580, Japan
[4] Graduate School of Pure and Applied Sciences, University of Tsukuba, 1-1-1 Tennodai, Tsukuba-shi, Ibaraki 305-8571, Japan
[5] Takasaki Institute for Advanced Quantum Science, National Institutes for Quantum Science and Technology (QST), Takasaki, Gunma 370-1292, Japan

*corresponding authors

**Corresponding author's e-mail address & phone number**
e-mail address: saito.hikaru.961@m.kyushu-u.ac.jp
Phone: 81-92-583-8829

**Abstract**

Precise allocation of nano light-sources in photonic integrated circuits is essential for the development of next-generation optical technologies such as optical nano-circuits, quantum information processing, and quantum communication. However, controlling the position of such light sources is a challenging task. We here show that halide perovskite nanoparticle light sources can be generated at the desired positions by electron beam. We applied cathodoluminescence spectroscopy to $CsPbBr_3$/$Cs_4PbBr_6$ composite produced by thermal evaporation and discovered that the intensity of green light emission from the $CsPbBr_3$ nanoparticles increased with electron beam irradiation. Changes in the cathodoluminescence spectrum associated with the electron beam irradiation suggest $CsPbBr_3$ nanoparticle formations. Furthermore, by taking advantage of the high spatial resolution and controllability of the electron beam, we demonstrate nano-light source patterning on the film.

Precisely arranging nanoscale light sources within nanophotonic structures enables advanced optical information processing, as demonstrated by observation of strong light-matter interaction in a single quantum dot coupled with a plasmon nanocavity[1], optical spin sorting by combination of quantum dots and topological photonic waveguides[2]. Ion or electron beams can potentially be used to fabricate nano light-sources with precise control of their positions since single photon emitters can be introduced by ion implantation[3–7]. Lattice defects introduced by ion or electron irradiation sometimes work as light sources[8–12]. By combining electron beam lithography and crystal growth on the semiconductor substrates, position-controlled growth of quantum dots have been realized[13–16]. In addition, top-down processes using in-situ electron beam lithography have been developed to control the coupling between on-chip optical circuits and introduced quantum dots[17]. Trapping excitons by introducing local strain in two-dimensional semiconductors using structured surfaces is also effective for deterministic positioning of single-photon emitters with nanocavities[18].

As light source material, halide perovskites have been attracting great attention, as they enable light emitting diodes with external quantum efficiencies exceeding 20%[19] and quantum dots functioning as single-photon emitters even at room temperature[20] with their luminescence wavelength being easily tuned[21]. However, there has so far been no established method for growing halide perovskite light sources locally on a substrate although integration of such nano-light sources into on-chip optical circuits is crucial for its practical use. In this study, we propose a new approach to generate nano light-sources by irradiating electron beams onto $CsPbBr_3$/$Cs_4PbBr_6$ composite, which exhibit high photoluminescence quantum yield[22]. To investigate the generated nano light-sources, we employed cathodoluminescence (CL), which utilizes fast electrons for nano-optics.

CsPbBr$_3$/Cs$_4$PbBr$_6$ composite powder were synthesized based on the protocol proposed by Chen et al[22]. Briefly, PbBr$_2$ (0.22 mmol, ≥ 98%, Sigma-Aldrich) and cesium acetate (0.88 mmol, 99.9% trace metals basis, Sigma-Aldrich) was stirred in dimethyl sulfoxide (0.5 ml, anhydrous, ≥ 99.9%, Sigma-Aldrich) at room temperature for 1 h. After adding 0.1 mL aqueous HBr solution (48% by weight in H$_2$O, Sigma-Aldrich), the mixed solution was stirred at room temperature for 12 h. The precipitation was separated from the mixed solution by centrifugation and the powder was dried under vacuum overnight. CsPbBr$_3$/Cs$_4$PbBr$_6$ composite nanocrystal films were deposited on a membrane substrate for transmission electron microscopy by thermally evaporating the synthesized precipitate: First, radio frequency (RF) magnetron sputtering was used to deposit a SiO$_2$ (10 nm) layer on the Si$_3$N$_4$ membrane grid (SiMPore, USA), and then the synthesized CsPbBr$_3$/Cs$_4$PbBr$_6$ powder was evaporated to form a 110 nm thick layer. Then, it was covered with a 10 nm SiO$_2$ layer by RF magnetron sputtering to prevent degradation due to exposure to air [Fig. 1(a)]. The fabricated film was analyzed by scanning transmission electron microscopy (STEM)-electron energy-loss spectroscopy (EELS). STEM-EELS was performed by using a transmission electron microscope Titan Cubed G2 60-300 (Thermo Fisher Scientific, USA) equipped with a monochromator and an energy filter Quantum 965 (Gatan, USA) operated at an acceleration voltage of 300 kV. Cathodoluminescence (CL) measurements were performed to investigate how the emission properties of this film change due to electron beam irradiation. STEM-CL was conducted using a homemade instrument built by modifying JEM-2100F (JEOL, Japan)[23], which was operated at an acceleration voltage of 80 kV. The probe diameter estimated from the illumination condition is 1~2 nm. This CL microscope is equipped with electrostatic dose modulator (JEOL, Japan), enabling instantaneous dose switching.[24] All measurements were performed at room temperature.

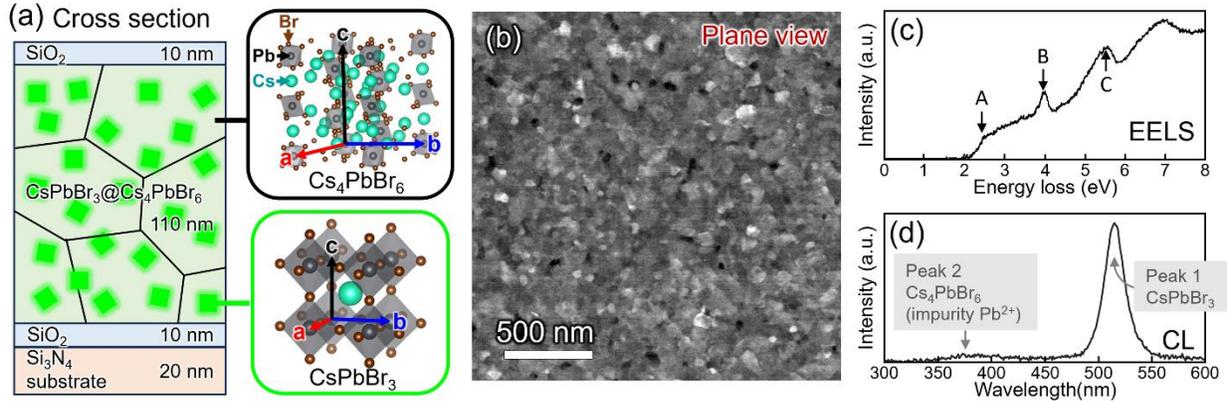

**FIG. 1.** STEM observation and EELS and CL analyses of the fabricated film. (a) Schematic crystal structures of $CsPbBr_3$ (cubic) and $Cs_4PbBr_6$ (rhombohedral), and a schematic multilayer structure of the film. (b) Plane view ADF-STEM image of the film. (c) EELS spectrum obtained by irradiating the entire field of view in (b), where the zero-loss peak was subtracted. (d) CL spectrum measured by irradiating the same film at a different field of view.

Fig. 1(b) shows a plane view annular dark-field (ADF) image of the fabricated multilayer film, suggesting that the deposited film is composed of fine grains of several tens of nanometers in size. The EELS spectrum [Fig. 1(c)], obtained by irradiating the entire ADF image area [Fig. 1(b)], shows the onset A and peaks B and C appearing below 6 eV, which is consistent with the previous report[25]. The onset A is attributed to an interband transition in the $CsPbBr_3$ phase and corresponds to its bandgap energy of ~2.3 eV[26–28]. The peaks B and C are consistent with those found by optical absorption spectroscopy of the $Cs_4PbBr_6$ phase[29,30]. These results confirm that the deposited film is a composite of $CsPbBr_3$ and $Cs_4PbBr_6$. Green light emission is observed as a peak at 515 nm in CL spectrum of this film ["Peak 1" in Fig. 1(d)], which is similar to the CL spectra previously obtained from $CsPbBr_3$ nanoparticles embedded in $Cs_4PbBr_6$ matrix ($CsPbBr_3@Cs_4PbBr_6$)[31]. The peak wavelength is shorter than that of bulk $CsPbBr_3$ (~530 nm)[32] similar to previous CL[31] and photoluminescence[22,33] spectra obtained from $CsPbBr_3@Cs_4PbBr_6$. The shortening of the emission wavelength with decreasing the particle size has been observed not only in bare $CsPbBr_3$ nanoparticles[34,35] but also in $CsPbBr_3@Cs_4PbBr_6$[31,33]. Since ~20 nm $CsPbBr_3$ nanoparticles in $Cs_4PbBr_6$ exhibit emissions in the wavelength range of 518 nm to 519 nm[22,33], the CL peak wavelength of 515 nm measured in this study suggests the average size of ~20 nm or less. Another peak is observed at a wavelength of ~375 nm (we call this Peak 2), which is known as emission from impurity $Pb^{2+}$ ions occupying the $Cs^+$ sites of the $Cs_4PbBr_6$ phase[36].

Surprisingly, when the same position is continuously irradiated with an electron beam, Peak 1 intensity increases as the irradiation time $t$ increases [Figs. 2(a) and 2(e)], and the peak height becomes more than doubled compared to the initial profile [Figs. 2(b) and 2(e)]. Assuming an exponential relaxation model $(I_0 - I_\infty)e^{-t/\tau} + I_\infty$, the relaxation time $\tau$ was deduced to be 13.1±1.5 s [solid red curve in Fig. 2(e)]. Here, $I_0$ and $I_\infty$ are the initial and final ($t \to \infty$) peak intensities, respectively. In contrast, Peak 2 becomes lower [Figs. 2(c), 2(d) and 2(f)] with irradiation time $t$, and its relaxation time is 5.2 ± 2.6 s [solid blue curve in Fig. 2(f)]. The intensity decrease of Peak 2 preceding the increase of the Peak 1 intensity suggests that $CsPbBr_3$ nanoparticles are formed (increase of Peak 1) after consuming the impurity $Pb^{2+}$ ions in the $Cs_4PbBr_6$ phase (decrease of Peak 2). The initial existence of Peak 2 indicates that the as-deposited $Cs_4PbBr_6$ phase initially has excess $Pb^{2+}$ ions. By the electron beam irradiation, the $Cs_4PbBr_6$ phase approaches the stoichiometric composition through the formation of the $CsPbBr_3$ phase. The relaxation time of Peak 1 is longer than that of Peak 2, suggesting crystallinity improvement proceeding for a longer period. However, excessive irradiation of the electron beam would rather introduce lattice defects in the $CsPbBr_3$ phase[36], which can lower the light emission.

The peak wavelength and width of Peak 1 hardly change during its intensity increase [Figs. 2(g) and 2(f)], suggesting that the $CsPbBr_3$ nanoparticles show almost no change in size, probably due to the lattice strain introduced by the lattice mismatch between $CsPbBr_3$ and $Cs_4PbBr_6$, which limits the crystal growth of $CsPbBr_3$[38]. Therefore, new $CsPbBr_3$ nanoparticles nucleate within the Cs$_4$PbBr$_6$ matrix by electron beam irradiation while the growth of the initially existing CsPbBr$_3$ nanoparticles is suppressed.

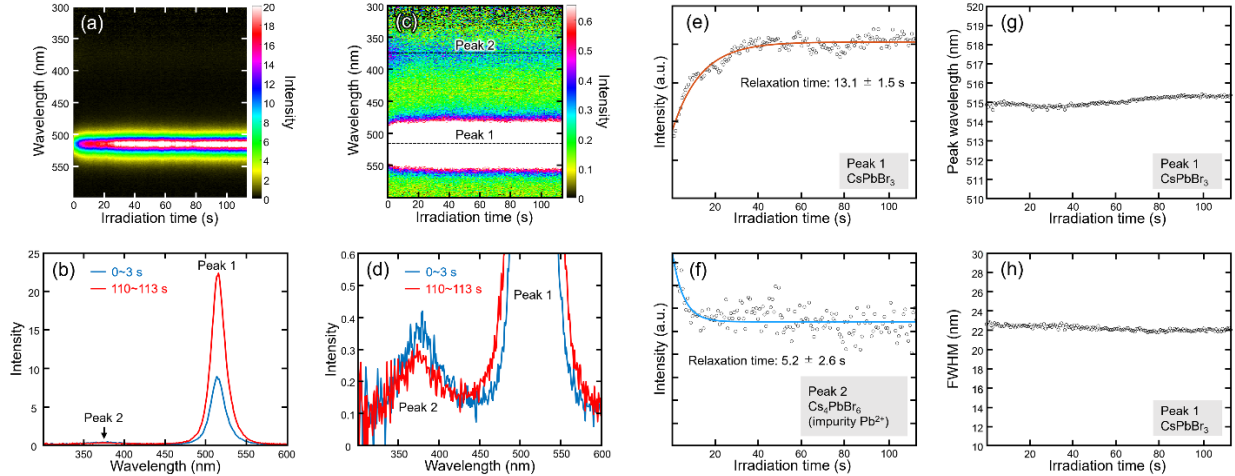

**FIG. 2.** Modification of the light emission property of the fabricated film by electron beam irradiation. (a) Irradiation time dependence of the CL spectrum obtained by setting the electron beam at the same spot with a probe current of 125 pA. The spectra were recorded every 0.74 seconds. (b) Time-averaged CL spectra at the early (0~3 s, blue profile) and late (110~113 s, red profile) stages of the continuous irradiation. (c,d) Intensity-magnified profiles of (a) and (b), respectively. (e–h) Irradiation time dependence of (e) height of Peak 1, (f) height of Peak 2, (g) wavelength of Peak 1, and (h) full width of half maximum of Peak 1, obtained by fitting each peak by a Gaussian function. The solid curve profiles in (e) and (f) are the fitting curves assuming exponential relaxation. The estimated relaxation times are $13.1 \pm 1.5$ s and $5.2 \pm 2.6$ s, respectively. The fitting errors are given based on 95% reliability.

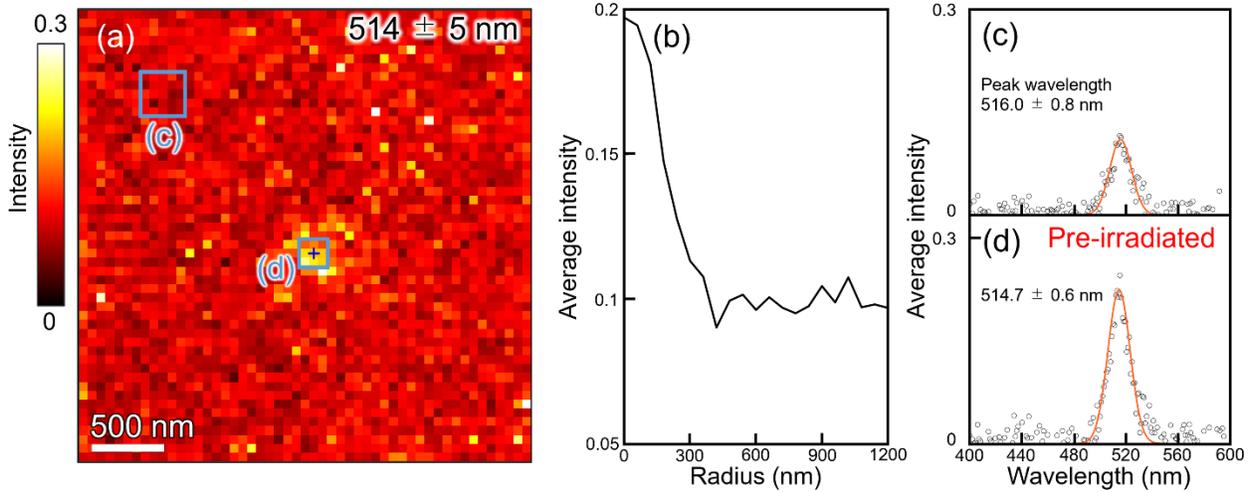

**FIG. 3.** CL mapping of an area changed by point irradiation. (a) CL intensity map averaged over the wavelength range of 514 ± 5 nm. Point irradiation was performed near the center of the field of view with a high probe current of 125 pA for 7 s prior to the mapping. CL mapping was performed with a low probe current of 1 pA, a scan step of 60 nm, and a dwell time of 0.3 s. This dose rate control was assisted by the electrostatic dose modulator. (b) Azimuthally averaged radial intensity profile around the bright center indicated by a blue cross in (a). (c, d) CL spectra extracted from the corresponding CL mapping positions indicated in (a), with the Gaussian fitting curves shown by red solid curves. The CL spectra are averaged over the square regions. The fitted peak wavelengths are displayed in each panel with the errors given by 95% reliability.

To evaluate the intensity-increased area by the electron beam irradiation, we pre-irradiated a spot by a high current electron beam (125 pA for 7 s) and performed CL mapping around this spot using a low probe current (1 pA, scan step of 60 nm, and a dwell time of 0.3 s) to avoid extra modification by the probe electron beam. Fig. 3(a) shows this CL intensity map averaged over the wavelength range of 514 ± 5 nm, corresponding to Peak 1. The pre-irradiated position, near the center of the map, exhibits a bright spot of submicron size. Fig. 3(b) shows the azimuthally averaged radial intensity profile around the bright center [blue cross in Fig. 3(a)]. The CL intensity increase spreads over a radius of approximately 300 nm, corresponding to the half width of ~300 nm in diameter. This modified range is more than one order of magnitude wider than the irradiation electron probe diameter of 1~2 nm, suggesting that long-range effect of sample heating by the electron beam is related to this modification. To investigate the spectral property of this area relative to the surroundings, we selected spectra from two representative positions, as shown in Figs. 3(c) and 3(d). The intensity of Peak 1 at the pre-irradiated position [Fig. 3(d)] is increased by approximately twice compared to the surroundings [Fig. 3(c)]. At the pre-irradiated position [Fig. 3(d)], the fitted spectrum shows a peak at $514.7 \pm 0.6$ nm, which is similar to the spectrum of surroundings [Figs. 3(c)], suggesting that a group of $CsPbBr_3$ nanoparticles of similar sizes were formed by the intense electron beam irradiation. Furthermore, a peak wavelength map shows that the CL peak wavelength exhibits no positional dependence across the entire field of view, including the pre-irradiated position (Supplementary Material).

To demonstrate the highly controllable positioning of the light-sources formed by electron beam irradiation, we drew a character string "FENO", standing for "fast-electron nano optics", with submicron line widths using a high probe current of 1 nA. The imprinted characters are visualized by CL mapping with a low probe current of 5 pA as shown in Fig. 4. The brighter areas are formed in the shape of the designed character string with submicron accuracy. This demonstration of drawing proves the potential of the proposed approach of placing nano light-sources at desired locations on nanophotonic devices.

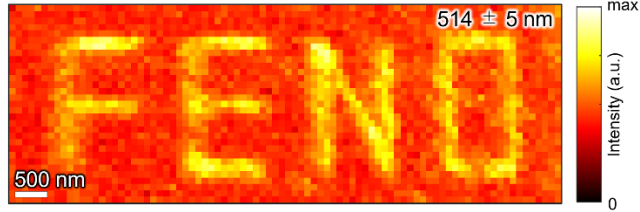

**FIG. 4.** Demonstration of light source drawing. The irradiation point of the 1 nA electron beam was successively varied along a letter "FENO" with a step of 50 nm and dwell time of 1 second. Then, a CL map was obtained with a low probe current of 5 pA, a scan step of 100 nm, and a dwell time of 0.6 s. This map displays the averaged intensity over the wavelength range of 514 ± 5 nm.

In summary, we discovered that the intensity of green light emission from the $CsPbBr_3$ nanoparticles increased with electron beam irradiation on a $CsPbBr_3$/$Cs_4PbBr_6$ composite film produced by thermal evaporation. By electron beam irradiation, $CsPbBr_3$ nanoparticles are formed by consuming the impurity $Pb^{2+}$ ions in the $Cs_4PbBr_6$ phase. The CL mapping further visualized the localization of the light emitter modified by the electron beam irradiation, which even allows drawing alphabets, demonstrating its applicability to the precise positioning of nano light-sources at desired positions. The proposed method has high compatibility with existing electron beam lithography and is expected to be incorporated into various nanophotonic device fabrications.

## CONFLICT OF INTEREST

The authors declare no competing financial interest.

## ACKNOWLEDGMENTS

This work was supported by JSPS KAKENHI Grant Number 25K01640, 23K17350, 22H05034, 22H01928.

[38] Z. Bao, Y.-J. Tseng, W. You, W. Zheng, X. Chen, S. Mahlik, A. Lazarowska, T. Lesniewski, M. Grinberg, C. Ma, W. Sun, W. Zhou, R.-S. Liu, and J. P. Attfield, *J. Phys. Chem. Lett.* **11**, 7637 (2020).15

# SUPPLEMENTARY MATERIAL

A. Peak height, wavelength, and width maps

B. Bright spots in surroundings

## A. Peak height, wavelength, and width maps

Fig. S1 shows spatial distributions of the height, wavelength, and full width of half maximum of the green CL peak ("Peak 1" in the main text) derived by Gaussian fitting applied to the CL spectral mapping data corresponding to Fig. 3. To ensure reasonable fitting accuracy, binning of 5 × 5 pixels was performed before the fittings. The results show that the intense electron beam irradiation only changes the peak height while the peak wavelength and width are almost the same as those of the surroundings.

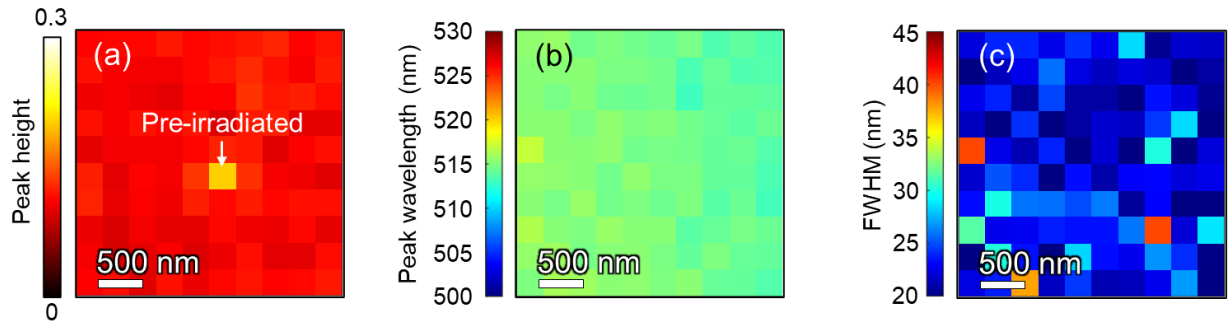

**FIG. S1.** Peak analysis of the CL spectral mapping data of Fig. 3 by Gaussian fitting. (a) Peak height, (b) wavelength, and (c) width maps of the green CL peak (Peak 1).

### B. Bright spots in surroundings

In Fig. 3(a) in the main text [shown in Fig. S2(a) again], single-pixel bright spots are scattered around the area away from the pre-irradiated position [e.g., (b) and (c) in Fig. S2(a)], and some are as bright as the pre-irradiated position. Figs. S2(b) and (c) show representative CL spectra of the bright spots. The peak wavelength is almost the same as the dark surroundings. This is due to sub-nanoscale heterogeneity in the spatial distribution of the $CsPbBr_3$ nanoparticles. These bright spots are all within a single pixel (= 60 nm) and can be distinguished from the intensity increased spot by the intense electron beam irradiation, which is larger than a single pixel (60 nm).

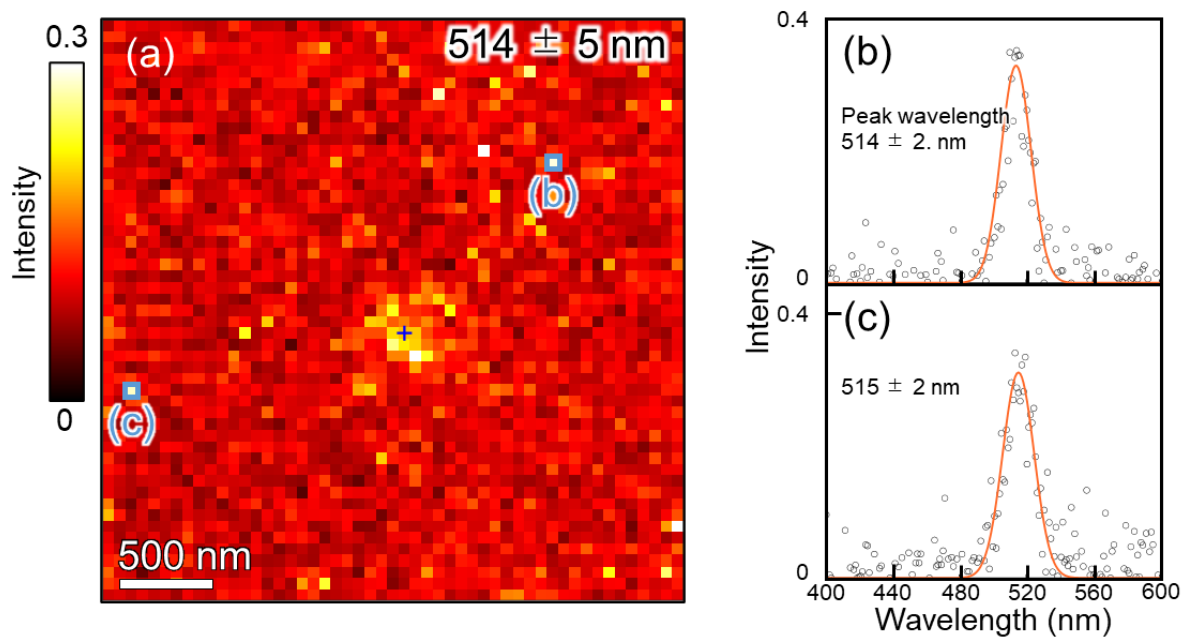

**FIG. S2.** Analysis of single-pixel bright spots in the CL map of Fig. 3(a). (a) CL intensity map averaged over the wavelength range of 514 ± 5 nm. (b, c) CL spectra (open circles) extracted from the corresponding CL mapping positions indicated in (a), with the Gaussian fitting curves shown by red curves. The fitted peak wavelengths are displayed in each panel with the errors given by 95% reliability.